\documentclass[preprint2]{aastex}
\usepackage{url}

\usepackage{graphicx} 

\RequirePackage{color}

\begin{document}

\title{Effect of data gaps on correlation dimension computed from light curves of variable stars}
%%\slugcomment{Not to appear in Nonlearned J., 45.}
%% Running heads
\shorttitle{Effect of data gaps-variable stars}
\shortauthors{Sandip V. George et al.}

\author{Sandip V. George}\and \author{G. Ambika}
\affil{Indian Institute of Science Education and Research, Dr Homi Bhabha Road, Pune-411008\\sandip.varkey@students.iiserpune.ac.in\\g.ambika@iiserpune.ac.in}
\and
\author{R. Misra}
\affil{Inter University Centre for Astronomy and Astrophysics, Ganeshkhind, Pune 411 007\\rmisra@iucaa.ernet.in}
%%\email{\emaila }

\begin{abstract}
Observational data, especially astrophysical data, is often limited by gaps in data that arises due to lack of observations for a variety of reasons. Such inadvertent gaps are usually smoothed over using interpolation techniques. However the smoothing techniques can introduce artificial effects, especially when non-linear analysis is undertaken. We investigate how gaps can affect the computed values of correlation dimension of the system, without using any interpolation. For this we introduce gaps artificially in synthetic data derived from standard chaotic systems, like the R{\"o}ssler and Lorenz, with frequency of occurrence and size of missing data drawn from two Gaussian distributions. Then we study the changes in correlation dimension with change in the distributions of position and size of gaps. We find that for a considerable range of mean gap frequency and size, the value of correlation dimension is not significantly affected, indicating that in such specific cases, the calculated values can still be reliable and acceptable. Thus our study introduces a method of checking the reliability of computed correlation dimension values by calculating the distribution of gaps with respect to its size and position.  This is illustrated for the data from light curves of three variable stars, R Scuti, U Monocerotis and SU Tauri. We also demonstrate how a cubic spline interpolation can cause a time series of Gaussian noise with missing data to be misinterpreted as being chaotic in origin. This is demonstrated for the non chaotic light curve of variable star SS Cygni, which gives a saturated D$_{2}$ value, when interpolated using a cubic spline. In addition we also find that a careful choice of binning, in addition to reducing noise, can help in shifting the gap distribution to the reliable range for D$_2$ values. 
\end{abstract}


%\section*{}
%\label{sec:intro}
\section{Introduction} 
\label{sec:Intro}
\noindent It is often of interest to establish and detect the nonlinearity underlying complex phenomena observed in nature. But in many such cases, one has to rely on observed data of one of the variables of the system to understand its underlying dynamics. For this, such data or time series are subjected to a number of tests, yielding a set of quantifiers. Such quantifiers or measures offer several indices to detect non trivial structures and characterize them effectively. One of the frequently used tests conducted on any time series to detect chaos is correlation dimension, D$_{2}$. Once the calculation of D$_{2}$ suggests the existence of chaos, other measures like Lyapunov exponents can be calculated for further confirmation. A popular algorithm used for calculating correlation dimension is the Grassberger-Procassia algorithm or GP algorithm\citep{Gra83}. According to this algorithm, the dynamics of a system is reconstructed, in an embedding space of dimension M, with the time series of a single variable using delay coordinates scanned at a prescribed time delay, $\tau$.  

Data from observations have several issues; primary among them being noise and uneven sampling and data gaps. Methods of analyzing chaos in signals from nonlinear systems contaminated with noise, has been a subject of extensive study previously\citep{Kos88}. While gaps are a major issue in observational data, its effects on chaos quantifiers have not been explored so far in detail. Non-uniform sampling can occur in a plethora of situations, such as in packet data traffic, astronomical time series, radars and automotive applications\citep{Eng07}. One of the primary kinds of non-uniform sampling is due to gaps in data, where the underlying data is uniformly sampled but  observations are frequently missed. Data from such kind of non-uniform sampling still has an inherent sampling frequency, of the underlying data. Observations can be missed for a number of reasons like human failure, instrument noise, imperfect sensors, unfavorable observational conditions etc. Various techniques, like interpolation, have generally been used to dress-up the data without considerable attention being given to the extent of their effects on the computed quantifiers. 

One of the areas where missing data is a major concern is in astrophysical observations. In optical astronomy, cloud cover, eclipses, unfavorable weather etc., are reasons for missing data. In radio astronomy, radio emissions from man made sources, ionospheric scintillation, lightning, emissions from the sun etc. are primary reasons to discard data samples. While a lot of these causes are resolved with the advent of space telescopes, gaps are found to occur even in data collected from these telescopes\citep{Gar14}. 

Data can also be totally unevenly sampled with no underlying sampling frequency. The study on variable stars from a dynamical systems view depends extensively on non linear analysis of observed data, requiring data over decades, almost daily for reasonable analysis. For observational data from variable stars, one has to rely mostly on data from amateur astronomers. Such data is highly prone to uneven sampling\citep{Buc95, Amb03} which makes standard methods of analysis impossible in such cases. Most data analysis in this context is conducted by using smoothing techniques to smooth over the unevenly sampled data\citep{Buc95}. The dangers of such smoothing techniques have been pointed out previously, in the context of climate data \citep{Gra86}. Methods to do correlation analysis, Fourier transforms etc have been reported for data with uneven or irregular sampling previously \citep{Reh11,Ede88,Sca89}. Binning through this data however can artificially introduce a binning frequency into the data. Thus the case of total uneven sampling can be reduced to a case of data with gaps. Difficulties related to computation of nonlinear measures and their reliability for such data are so far not studied in detail.  

%%Such data occur in geophysical context, like paleo-climatic proxy data, and even in certain astrophysical contexts\citep{Gro97, Ste94, Buc95}.%%
 
In the present paper we propose to conduct an extensive and exhaustive analysis of the effect of gaps on nonlinear data analysis. The importance of correlation dimension as an indicator of chaos is exemplified by the wide range of fields where it has been used successfully. These include analyzing EEG data of the human brain, stellar pulsations, gearbox condition monitoring, stock market data etc.\citep{Fre85, Buc95, Jia99, Oh02}. While a non integer saturating correlation dimension is usually indicative of chaos in the system, it may be noted that there are a certain class of deterministic non linear systems which are strange and non-chaotic, i.e. they exhibit fractal nature but do not exhibit chaos\citep{Hea91}. Recently, variable stars that exhibit strange non-chaotic behavior, was identified using data from the Kepler space telescope\citep{Lin15}. The effects of noise and insufficient data on D$_{2}$ calculations have been studied previously\citep{Har06, Har09, Rue90, Lor91, Tso88}. Here we propose to study the effect of missing data on computed D$_2$ values. For this we consider data from standard nonlinear systems like the Lorenz and R{\"o}ssler systems, and artificially introduce gaps into them. Such artificially generated data sets are then subject to studies of correlation dimension as functions of the size and the frequency of data removal. As test cases we also subject two of the artificially generated time series to surrogate analysis to lend confidence to our results. We test the results of our analysis to real world systems, by applying them to unevenly sampled variable star data.  For this we do a suitably chosen binning to change the case of pure uneven sampling to one of data with gaps.  If the bin size is carefully chosen, this may also help to shift the gap distribution to give reliable estimates for D$_{2}$.  From the results of our analysis we define an acceptable range within which gaps in data would not considerably affect the calculation of D$_{2}$ from observational data. This would prove useful to scientists trying to detect and quantify chaos in systems based on such time series. We also illustrate the drawbacks associated with interpolation done on a typical time series with gaps.  We introduce gaps in a time series of random numbers drawn from a Gaussian distribution, and show that interpolation can cause the random data to seem to have some underlying nonlinear dynamics. This is also illustrated for the case of the non chaotic light curve of variable star SS Cygni. 

\section{Correlation dimension analysis of synthetic data with gaps} 
\label{sec:analysis} 
The starting point in our analysis is the adaptation of the standard GP algorithm, by \citep{Har06}, for calculating the correlation dimension in a non-subjective manner. The algorithm first converts a data set into a uniform deviate. The time delay $\tau$ is chosen as the time at which the auto correlation falls to $\frac{1}{e}$. For each embedding dimension, M, correlation dimension $D_{2}$(M) 
is computed. The resultant D$_{2}$(M) curve is fitted using the function \citep{Har06} 
\begin{eqnarray} 
f(M)\; & = & \;\Big {(}{D_2^{sat} - 1 \over M_d -1}\Big {)} (M-1) +1 
\;\;\;\; \hbox {for} \;\;\;\; M < M_d \nonumber \\ 
          & = & \; D_2^{sat} \;\;\;\; \hbox {for} \;\;\;\; M 
\geq M_d 
\label{fit} 
\end{eqnarray} 
which returns the saturated correlation dimension D$_{2}^{sat}$ with an error estimate (In this paper "correlation dimension" will refer to this "saturated correlation dimension" unless explicitly stated otherwise). Here M$_d$ is the embedding dimension, at which the D$_2$ vs M curve saturates. It represents the minimum dimensions of phase space required for reconstructing the dynamics of the system. We use this algorithm to calculate the correlation dimension of two standard dynamical systems, the R{\"o}ssler and the Lorenz systems and then the variation in D$_{2}$ in these systems when gaps are introduced artificially. 

\begin{figure}[h!]
\centering 
\plotone{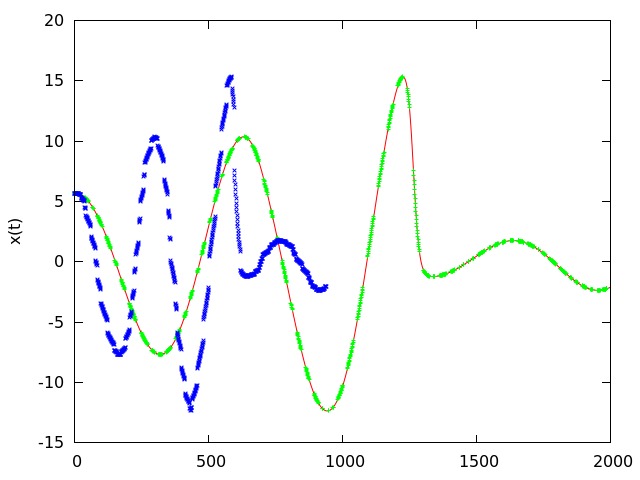} 
\caption{\label{fig:time_series} Time series of R{\"o}ssler system with and without gaps. The full curve(in red) represents the time series without gaps, the points in green show the time-series after introduction of gaps and points in blue shows the resultant time-series after merging the data, ignoring the gaps.} 
\end{figure}

We observe that, in real systems, gaps occur at random throughout the data set, and the size of gaps or missing data too is random. Since such gaps arise from multiple independent sources, it seems reasonable to assume that both the position and size of the gaps follow Gaussian distributions that can be characterized by their specific mean, m and standard deviation $\omega$. Starting with large gapless data-sets(of around 10$^{6}$ points) from the x variable of two standard systems, we use two Gaussians for removing data, G$_S$ with mean m$_{s}$ and deviation $\omega_s$, decides  the size of data removed and G$_P$ with mean m$_p$ and deviation $\omega_p$,  is used to locate the positions to remove data. The former models the size of the gaps and the latter models the width of regions in between where there are no gaps. Thus G$_P$ is an inverse measure of frequency of gaps since having a small value for m$_p$ will mean more frequent gaps.  The distribution of gaps can be characterized uniquely using these two Gaussian distributions given by  
\begin{equation} 
G_S(s;m_s,\omega_s) = \frac{1}{{\omega_s \sqrt {2\pi } }}e^{{{ - \left( {s - m_s } \right)^2 } \mathord{\left/ {\vphantom {{ - \left( {x - m_s } \right)^2 } {2\omega_s ^2 }}} \right. \kern-\nulldelimiterspace} {2\omega_s ^2 }}} 
\end{equation} 
\begin{equation} 
G_P(p;m_p,\omega_p) = \frac{1}{{\omega_p \sqrt {2\pi } }}e^{{{ - \left( {p - m_p } \right)^2 } \mathord{\left/ {\vphantom {{ - \left( {x - m_p } \right)^2 } {2\omega_p ^2 }}} \right. \kern-\nulldelimiterspace} {2\omega_p ^2 }}} 
\end{equation} 
The data removal is done such that the resulting data-sets all have $\approx$ 20000 points each. The sampling time of the original gapless dataset is chosen such that the reconstructed attractor should cover all its typical parts using the 20000 points. Thus the sampling time of the R{\"o}ssler is taken to be 0.1 and for the Lorenz is taken to be 0.01. After removal of data, the time series is condensed, ignoring the gaps. A section of time series data thus obtained for the x variable of R{\"o}ssler system before and after removal of data is shown in Fig.\ref{fig:time_series}.  
We first calculate the correlation dimension for the gapless datasets from the R{\"o}ssler and Lorenz systems, which we refer to as D$_2^0$. Then the data with gaps is analyzed for variation of correlation dimension D$_{2}$. The natural time scale in the GP algorithm is the delay time, $\tau$, that is used to construct the pseudo-vectors. Hence m$_{s}$  and m$_{p}$ of gap distributions are considered in units of $\tau$. This helps to minimize the dependence on sampling frequency, as the gaps are considered in units of $\tau$ which is invariant with sampling frequency. As expected the autocorrelation falls faster when we have data with gaps. The vectors are constructed using this smaller value of $\tau$. However, it is noticed that the variation of $\tau$ in the ranges considered is not very high. 

\begin{figure}[h!]
\centering 
\plotone{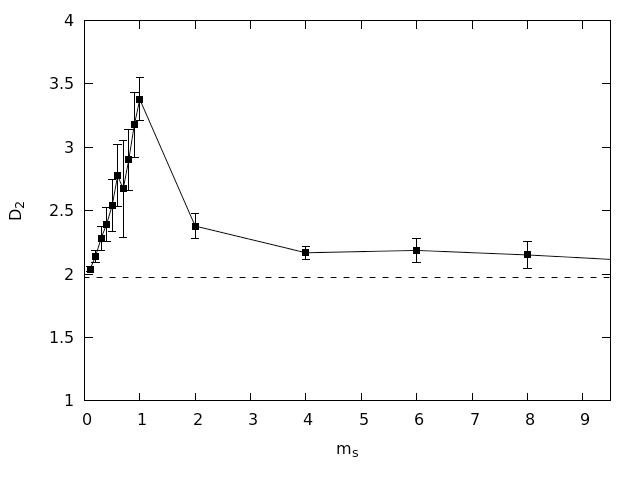}
\plotone{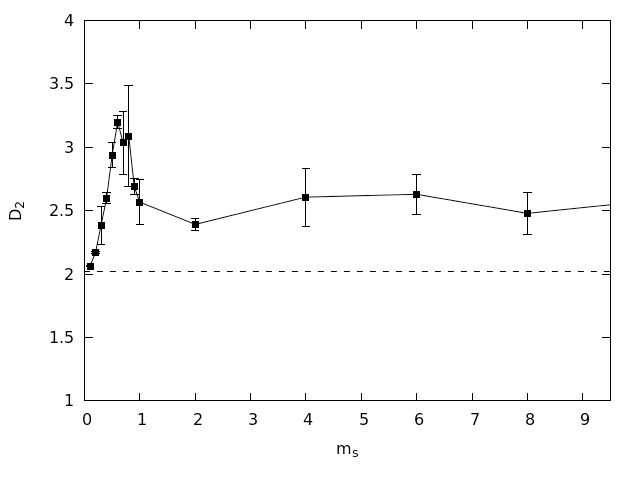}
\caption{\label{fig:var_size} Correlation dimension from (a)Lorenz and (b)R{\"o}ssler data with changing mean gap size, m$_{s}$, for a fixed m$_p$=9$\tau$. The x axis is in units of delay time, $\tau$. There is considerable deviation from the evenly sampled value $\sim$ 2 around 1$\tau$. The error-bars represent the standard deviation of 5 different realizations. The dotted line parallel to the x-axis shows the value of D$_2^0$.} 
\end{figure}

\begin{figure}[h!]
\plotone{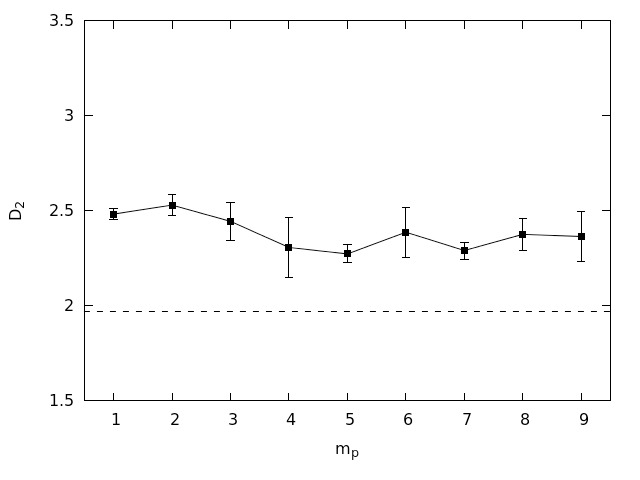}
\plotone{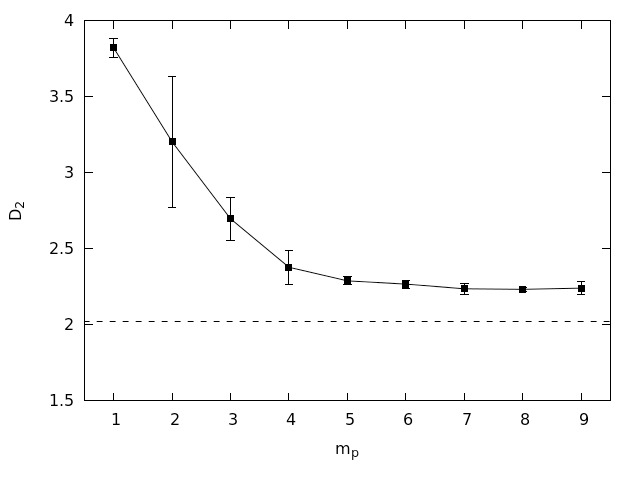} 
\caption{\label{fig:var_pos} Correlation dimension from (a)Lorenz and (b)R{\"o}ssler data with changing mean gap position, m$_{p}$, for a fixed gap size m$_{s}$=2$\tau$. The x axis is in units of delay time, $\tau$. The D$_{2}$ value of the Lorenz data is less affected by highly frequent gaps, when compared to the R{\"o}ssler data. The error-bars represent the standard deviation of 5 different realizations. The dotted line parallel to the x-axis shows the value of D$_2^0$.} 
\end{figure}

\begin{figure}[h!]
\centering 
\plotone{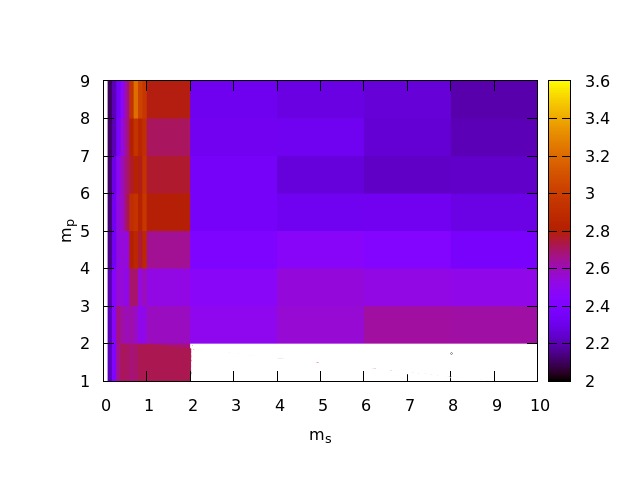} 
\plotone{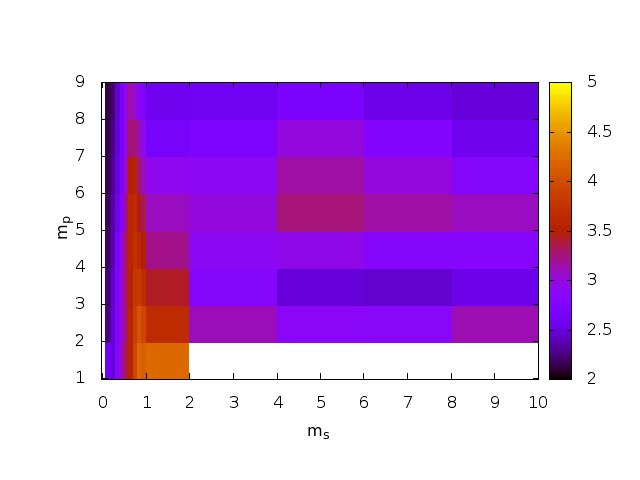} 
\caption{\label{fig:par_plane} Correlation dimension from (a)Lorenz and (b)R{\"o}ssler data with changing mean gap size, m$_{s}$, and mean gap position m$_{p}$. The axes are in units of delay time, $\tau$, and the color code is based on values of D$_2$. The white region corresponds to no saturation. The blue regions are closer to the actual value of D$_{2} \sim$ 2. The deviation from this value is visible in both graphs at m$_{s} \sim$ 1$\tau$ across different values of m$_p$. The standard deviation, $\omega_{p}$ is fixed at 0.1m$_p$ and $\omega_{s}$ at  0.1m$_s$. } 
\end{figure}

\begin{figure}[h!] 
\plotone{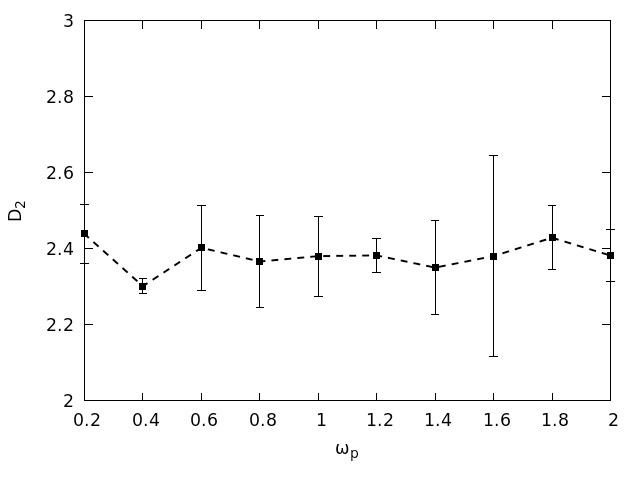} 
\plotone{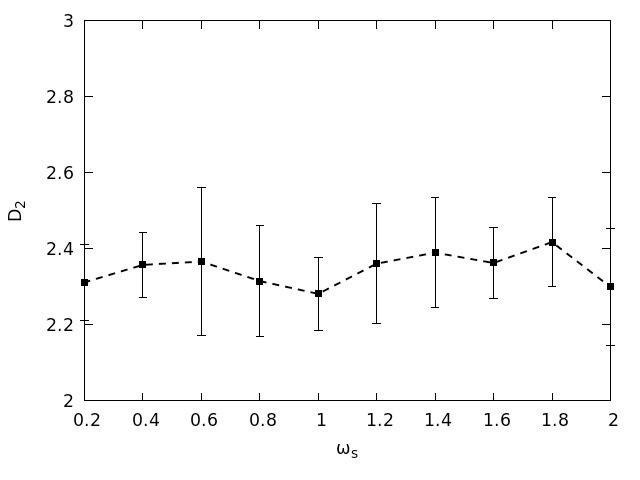} 
\caption{\label{fig:var_dev} Correlation dimension from Lorenz data with changing standard deviation of position, $\omega_{p}$ and $\omega_{s}$.The x axis is in fraction of mean position, m$_{p}$ and mean size m$_{s}$ respectively. m$_{p}$=9$\tau$ and m$_{s}$=9$\tau$ for both figures. The variation for D$_2$ with varying standard deviations is not significant in both cases.} 
\end{figure}

We first look at the variation of correlation dimension with change in mean of gap size, m$_{s}$, when the mean of gap position(m$_{p}$) is kept fixed. For this, the standard deviations, $\omega_{s}$ and $\omega_{p}$ are fixed at 0.1 m$_{s}$ and 0.1 m$_{p}$ respectively. We find that the D$_{2}$ value increases with increasing mean size, m$_{s}$, reaches a maximum at around m$_{s}$ $=$ $\tau$, and falls, at higher values of m$_{s}$. A representative case is shown for m$_{p}$ $=$ $9\tau$ in Fig. \ref{fig:var_size} for the Lorenz and R{\"o}ssler data. The variation of correlation dimension is then studied for change in m$_{p}$, keeping m$_{s}$ constant. In this case, the D$_{2}$ value decreases with  increasing value of m$_{p}$, reaching the value, D$_{2}^0$ for large m$_{p}$.  A representative case is shown for m$_{s}$ $=$ $2\tau$ in Fig. \ref{fig:var_pos} for both Lorenz and R{\"o}ssler data. It is noticed that the variation with frequency is higher for the R{\"o}ssler  than the Lorenz. The variations in the computed values of D$_2$, for the Lorenz and R{\"o}ssler data  across different values of m$_s$ and m$_p$, is shown in Fig. \ref{fig:par_plane}. We study the variation of D$_{2}$ with standard deviations, $\omega_{s}$ and $\omega_{p}$, with the means, m$_{s}$ and m$_{p}$, kept fixed. In this case, there seems to be no significant change in D$_{2}$ with change in standard deviation of either. A typical graph is shown in Fig.\ref{fig:var_dev} for Lorenz data, where the $\omega_{s}$ and $\omega_{p}$ are taken as fractions of mean size, m$_s$ and mean position, m$_p$.



Another interesting point to consider would be the value of the embedding dimension at which D$_{2}$ saturates, M$_{d}$. We find as the parameters of the introduced gaps vary, M$_{d}$ typically varies between 3 and 6, for both Lorenz and R{\"o}ssler data(where 3 should be the embedding dimension of both systems with evenly sampled data). However this happens when D$_2^{sat}$ too deviates highly from the evenly sampled case. Thus even when the correlation dimension saturates, the higher value of M$_d$ obtained from a data series with gaps can lead to another erroneous conclusion that the system is of higher dimensions than it actually is. 

\begin{figure}[h!] 
\plotone{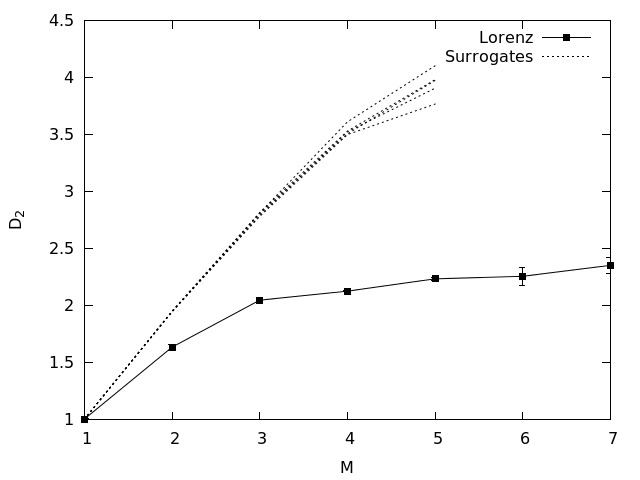} 
\plotone{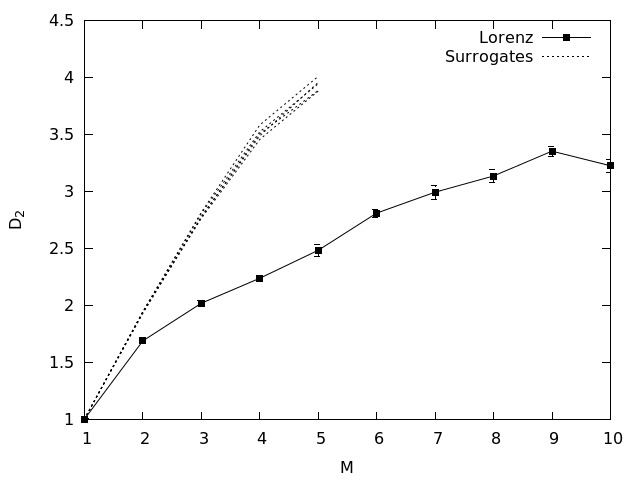} 
\caption{\label{fig:surr} Plot of D$_{2}$ vs M for the case of Lorenz data with gaps and its surrogates. (a) m$_{s}$=10$\tau$, m$_{p}$=9$\tau$  and (b) m$_{s}$=0.8$\tau$. m$_{p}$=9$\tau$. In both cases the surrogates deviate significantly from the actual data, showing that the data behaves differently from noise.} 
\end{figure}

We also subject two of the artificially generated time series to surrogate analysis. The surrogates are generated using the method of Iterative Amplitude Adjusted Fourier Transform (IAAFT) \citep{Sch96} implemented in TISEAN package\citep{Heg99}. This method is applicable only for cases with even sampling. Hence we generate five surrogate data-sets for the evenly sampled data, before removing data. Gaps are then introduced into the data as described before. Similarly, gaps are introduced into the surrogates with the same series of random numbers as used for data removal in the original dataset. These sets of data and surrogates with gaps are then subject to correlation dimension analysis. We take the cases where the D$_{2}$ vs M curve saturates close to the evenly sampled value and another where the D$_{2}$ vs M curve saturates with a value away from the evenly sampled value. All D$_{2}$ vs M curves for the surrogates in both cases are very different from the curve for the original data set as shown in Fig.\ref{fig:surr}. This would mean that even though gaps in data can result in the D$_2$ values deviating away from the original value in certain cases, it still behaves differently from noisy data. 

\begin{figure}[h!] 
\centering 
\plotone{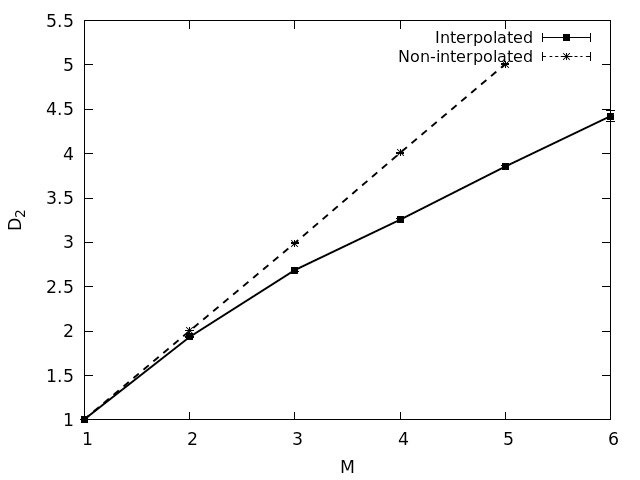} 
\caption{\label{fig:intvsno} Plot of D$_{2}$ vs M for a time series of random numbers with gaps, with(full curve) and without interpolation(dotted curve). The time-series does not saturate for the non interpolated case, while it saturates for the interpolated case.} 
\end{figure} 

Finally we consider the effect of smoothing data through interpolation. For this purpose, we start with a time series of random numbers drawn from a Gaussian distribution and artificially introduce gaps with m$_{p}$ of 300 data points, and varying m$_{s}$ from 3 to 350 data points. Subsequently, all the data sets are interpolated through, using a cubic spline and subject to calculation of D$_{2}$. For the non interpolated case, none of the data sets saturate. For the interpolated data sets, when the gap size is larger than 250 data points, we see that the D$_{2}$ curves saturate, yielding a value for correlation dimension, and hence suggesting an underlying nonlinearity in the system(Fig.\ref{fig:intvsno}). 

Thus our analysis yields two regions where gaps in data seems to affect correlation dimension. The first among these is in the low m$_{p}$  region. This region is more prominent in the R{\"o}ssler data, where the D$_{2}$ value comes within 1.3 times D$_2^0$, only when the mean separation between gaps is about 4$\tau$. In the Lorenz data this effect is not that pronounced. The second critical region occurs in the region of mean gap size, centered around 1$\tau$. Here we see, in both data at around this region, the D$_2$ value increases to above 1.5 times D$_2^0$. The D$_2$ value remains acceptable($<$1.3*D$_2^0$) in the remaining region of m$_s$ and m$_p$. We also find that varying the values of standard deviations, $\omega_s$ and $\omega_p$ do not affect the value of D$_2$ too much.

\section{Application to variable star data} 
\label{sec:App} 
In this section, we present the application of the analysis given in the previous section to cases of real world observational data sets. In this context  we note that astrophysical datasets are highly prone to gaps, as unlike controlled laboratory observations astrophysical observations are limited by atmospheric conditions, noise contamination etc. We consider the light curves of four irregular variable stars, R Scuti (R Sct), U Monocerotis (U Mon), SU Tauri (SU Tau) and SS Cygni (SS Cyg). Of these R Sct and U Mon are RV Tauri stars of RVa and RVb subcategories respectively and SU Tau is an RCB class variable star\citep{Och00,Per07}. All three stars have previously been either identified or suspected to have non linear dynamics\citep{Mal10, Buc95, Kol90, Dic91}. SS Cygni however is a cataclysmic variable star, of the dwarf nova class, established as not having any non linear properties\citep{Can88,Can92}. 

We analyze the light curves of all the four stars from 1950 to 2014 available on the American Association of Variable Star Observers (AAVSO) database. This data is highly unevenly sampled. As mentioned in Section~\ref{sec:Intro}, purely unevenly sampled data or irregular sampling offers a challenging situation\citep{Gro97,Hu07,Her06}. The distribution of time intervals, $\Delta$t, is what is considered in such cases. This $\Delta$t distribution is often treated as a Gamma distribution\citep{Reh11}. The limiting case of very small m$_p$ removes the underlying sampling frequency. In the limit of small m$_s$ and m$_p$ the generated data with gaps can be shown to approximate to a case of pure uneven sampling with Gamma distribution for $\Delta$t. As mentioned before, in Section~\ref{sec:Intro}, binning through pure unevenly sampled data can introduce an artificial sampling time, corresponding to the bin size. This converts the case of pure uneven sampling to a case of data with gaps. Hence we bin the light curves of the variable stars under consideration suitably, before analysis. For binning, we average through all the values within the chosen bin size “b”. While doing this sequentially, if the total number of points within b is less than a small number n, we leave it as a gap; otherwise replace it by the average of all values within the bin.  In our calculations we have taken n to be 3. In addition to convertion of unevenly sampled data to one with gaps, we also reduce the noise present in the data through binning.

After binning the variable star data using appropriate time intervals of two or five days, we first find the mean gap size and mean gap position for each and then decide whether the calculation of D$_{2}$ will yield an acceptable value, according to our earlier analysis. This is checked by direct computation of the D$_{2}$ values. 
\subsection {Variable star R Scuti}
\begin{figure}[h!]
\centering
\plotone{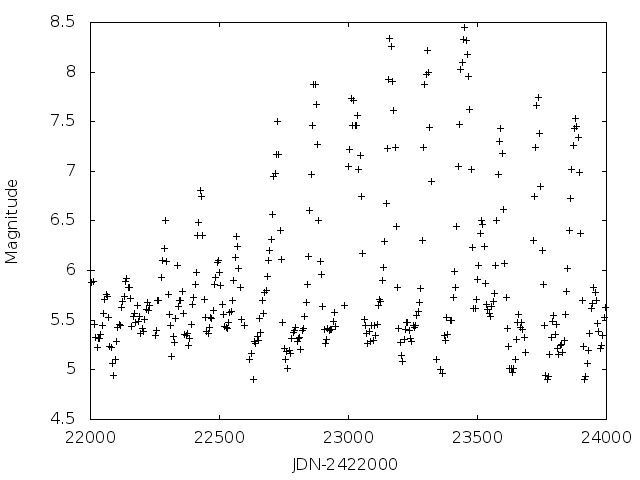}
\caption{ Light curve of variable star R Scuti binned to 5 days.}\label{fig:rsctlc}
\end{figure}
\begin{figure}[h!]
\centering
	\plotone{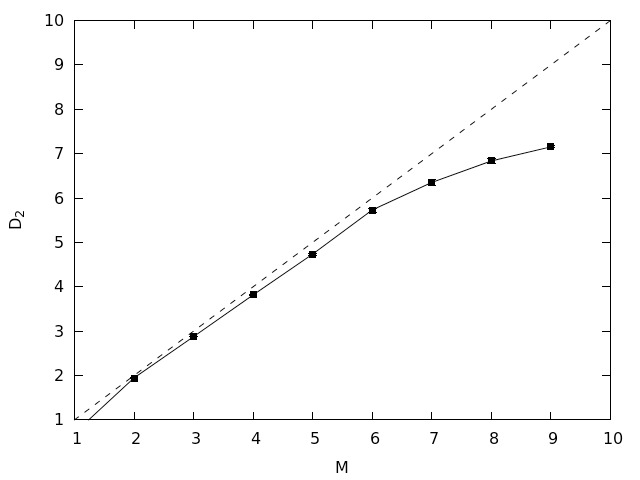}
	\plotone{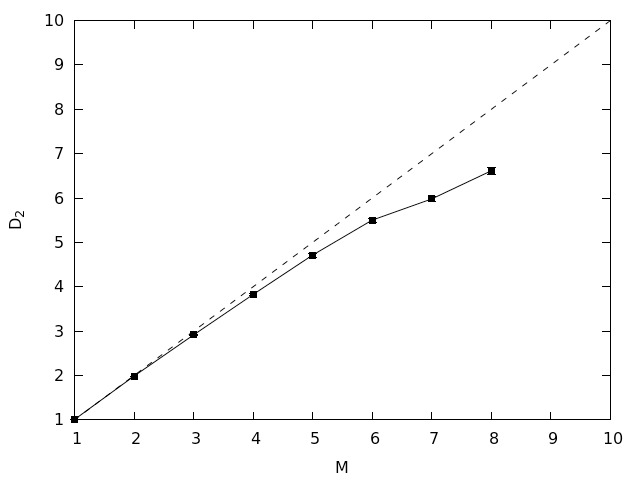}
	\plotone{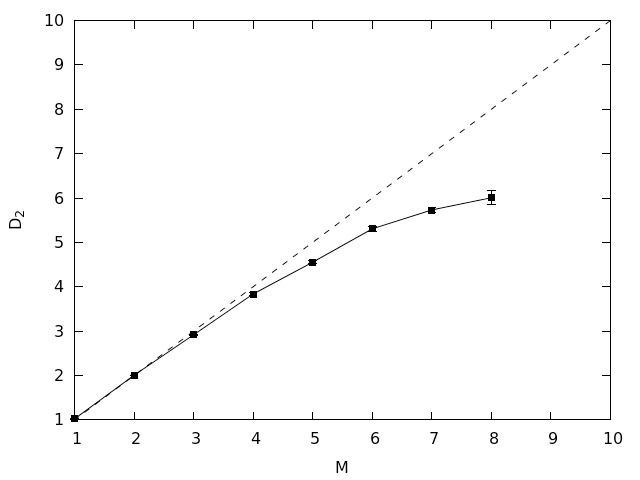}
\caption{ Correlation dimension D$_{2}$ vs embedding dimension, M, for variable star R Scuti. The curves do not saturate for cases (a) and (b)binned over 2 and 5 days respectively, whereas it saturates for case (c)binned over 10 days. The dotted curve shows the non saturating case corresponding to white noise.}\label{fig:rsct}
\end{figure}

The delay time of the AAVSO data-set for R Scuti is found to be approximately 20 days. After binning to 2 days, we find the distribution of gaps in this data-set has mean of the gap size distribution, m$_{s}$ $\sim$ 8 days and mean position, m$_{p}$ $\sim$ 33 days. In terms of the delay time $\tau$, m$_{p}$, is 1.65$\tau$ and m$_{s}$ is 0.4$\tau$. While the size falls outside the critical range for our analysis, the high frequency casts questions on the calculation of value of correlation dimension. The D$_{2}$ vs M graph does not saturate as feared, and no value for D$_{2}^{sat}$ can be determined. We again bin the data set to 5 days. A section of this light curve is shown in Fig.\ref{fig:rsctlc}. The value of m$_{p}$ now goes up to $\approx$ 10.6$\tau$, whereas m$_{s}$ goes up to $\approx$ 1.1$\tau$. Despite the higher value of m$_{p}$, m$_{s}$ falls into the critical range identified, and as expected, no saturated correlation dimension could be obtained, using our prescription for saturation mentioned in Section 1. Finally, we attempt to bin the data set over 10 days, yielding an m$_{p}$ value of $\approx$ 16.4$\tau$ and  and an m$_{s}$ value of $\approx$ 1.4$\tau$. Again the m$_{s}$ value is close to the critical region identified. The D$_{2}$ now shows a tendency to saturate to a value of 5.67$\pm$0.10. We would tend to doubt this value because the mean gap size is close to the critical region. We note that this value is significantly higher than D$_{2}$ values reported earlier using interpolated data-sets from AAVSO\citep{Buc95,Kol90}. The plot of correlation dimension D$_{2}$ vs embedding dimension, M for the three cases is shown in Fig.\ref{fig:rsct}.
\subsection{Variable Star U Monocerotis}
\begin{figure}[h!]
\plotone{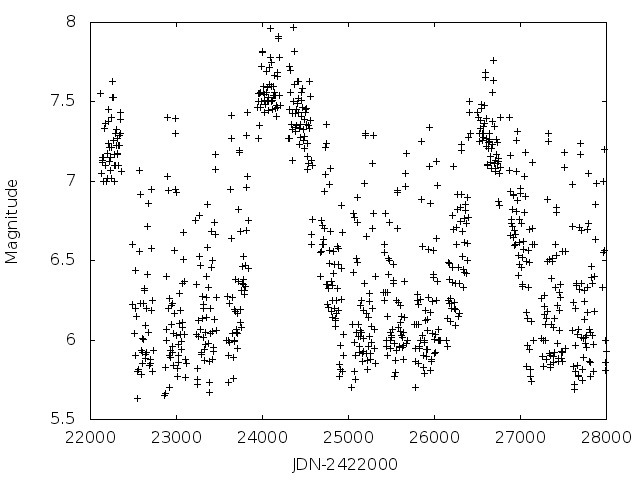}
\caption{ Light curve of variable star U Monocerotis binned over 5 days.}\label{fig:umonlc}
\end{figure}

\begin{figure}[h!]
		\plotone{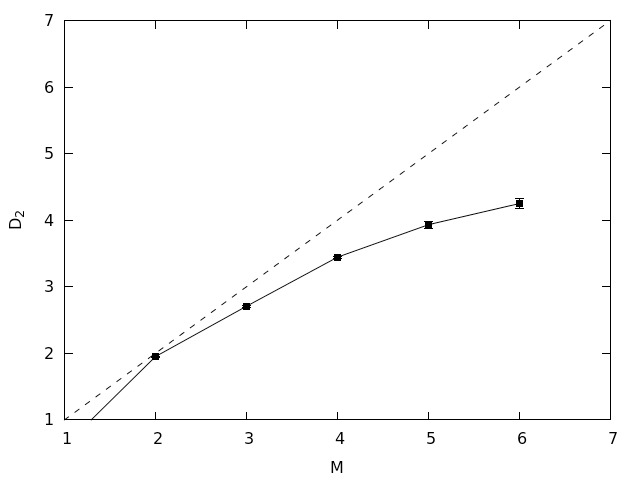}
		\plotone{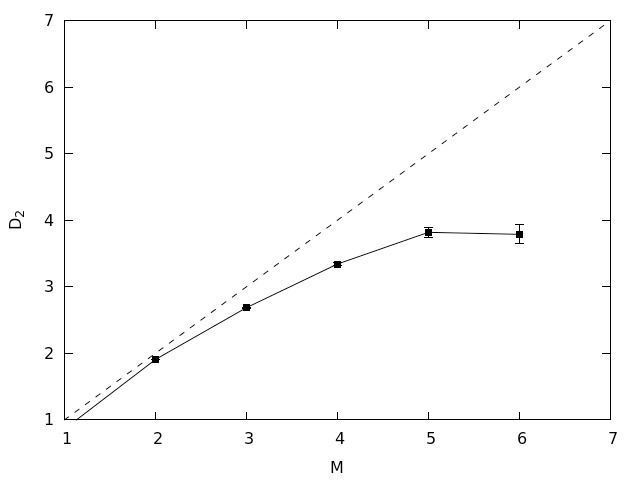}
\caption{ Correlation dimension D$_{2}$ vs embedding dimension, M, for variable star U Monocerotis. Using equation 1, The curve does not saturate for case (a) binned over 2 days, while it saturates for case (b) binned over 5 days. The dotted curve shows the non saturating case corresponding to white noise.}\label{fig:umon}
\end{figure}

We do a similar analysis for data from variable star U Mon, of the RVb subcategory of RV Tauri variables\citep{Och00}. A section of this light curve from AAVSO for 64 years from 1950 to 2014 is shown in Fig.\ref{fig:umonlc}. The delay time, $\tau$, for this data set is found to be $\approx$15 days. For bin size of 2 days, m$_{p}$ is found to be 11.4 days $\approx$ 0.76$\tau$. The mean size, m$_{s}$ of the data set is found to be 15.87 days $\approx$ 1$\tau$. Both m$_{p}$ and m$_{s}$ fall into critical ranges for calculation of D$_{2}^{sat}$, and as expected, the D$_{2}$ vs M graph\ does not saturate. We next bin the data-set into 5 day bins. This yields m$_{p}$, $\approx$ 3.5$\tau$ and m$_{s}$ $\approx$ 2.9$\tau$. Both m$_{p}$ and m$_{s}$ fall outside the critical ranges for calculation of D$_{2}^{sat}$. The D$_{2}$ vs M graph saturates, yielding a saturated correlation dimension of 3.44$\pm$0.08. To the best of our knowledge no previous reported value exists for comparison in this case. However based on our analysis with standard data sets, we expect this to be a reliable value. The plot of correlation dimension D$_{2}$ vs embedding dimension, M is shown in Fig.\ref{fig:umon}.
\subsection {Variable Star SU Tauri}
\begin{figure}[h!]
\centering
\plotone{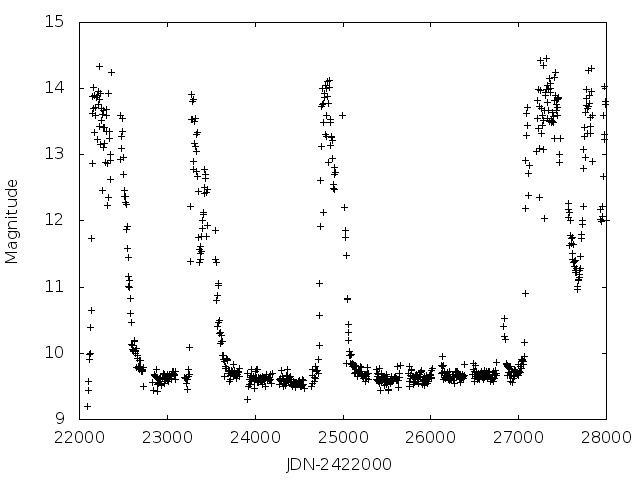}
\caption{ Light curve of variable star SU Tauri binned over 5 days.}\label{fig:sutaulc}
\end{figure}
\begin{figure}[h!]
\centering
	\plotone{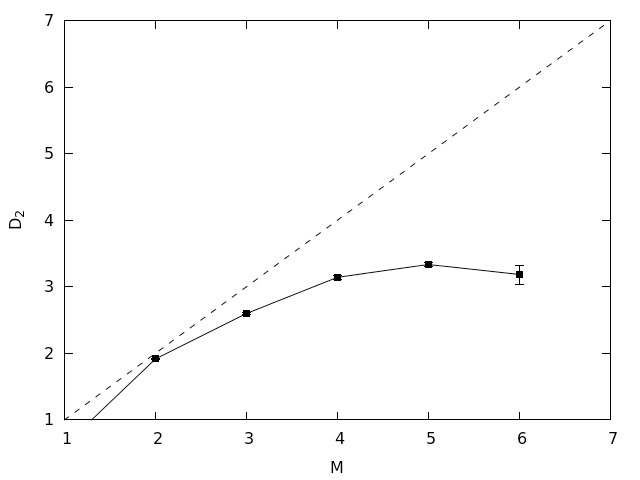}
	\plotone{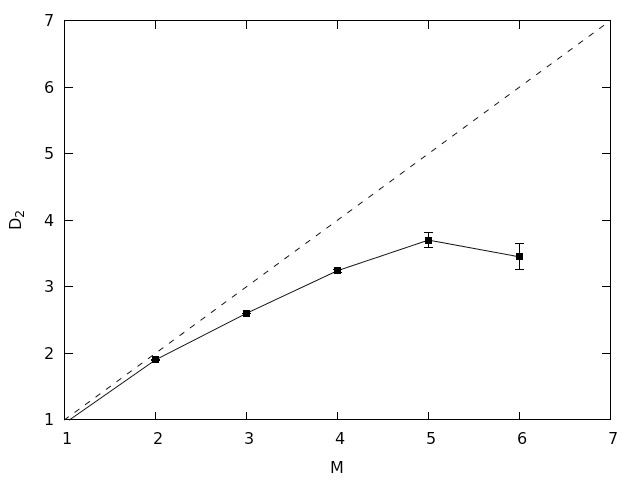}
\caption{ Correlation dimension D$_{2}$ vs embedding dimension, M, for variable star SU Tauri. Both cases (a) and (b), binned over 2 and 5 days respectively, saturate. The dotted curve shows the non saturating case corresponding to white noise.}\label{fig:sutau}
\end{figure}

A similar analysis is done using the data from variable star SU Tau, of the R Coronae Borealis type\citep{Och00}. The chaotic nature of the brightness variation for this star was suspected previously \citep{Dic91}. The AAVSO data set for this star is from 1950 to 2014 and a section of this light curve is shown in Fig.\ref{fig:sutaulc}. The delay time,$\tau$, for this data set is found to be 224 days. On binning to 2 days, the value of m$_{p}$ $\approx$ 0.06$\tau$ and m$_{s}$ $\approx$ 0.06$\tau$. Whereas the frequency itself is quite high, the gap size is well outside the critical ranges. We find the D$_{2}$ vs M graph saturates and yields a value of D$_{2}^{sat}$ of 3.26$\pm$0.08. With binning to 5 days, m$_{p}$ changes $\approx$ 0.28$\tau$ and m$_{s}$  $\approx$ 0.15$\tau$. We find m$_{s}$ is still well outside the critical range and the D$_{2}$ vs M graph saturates in this case also, yielding a value of D$_{2}^{sat}$ of 3.28$\pm$0.05, consistent with the previous value. In this case also, there is no reported value for comparison. The plot of correlation dimension D$_{2}$ vs embedding dimension, M is shown in Fig.\ref{fig:sutau}.
\begin{figure}[h!]
\centering
\plotone{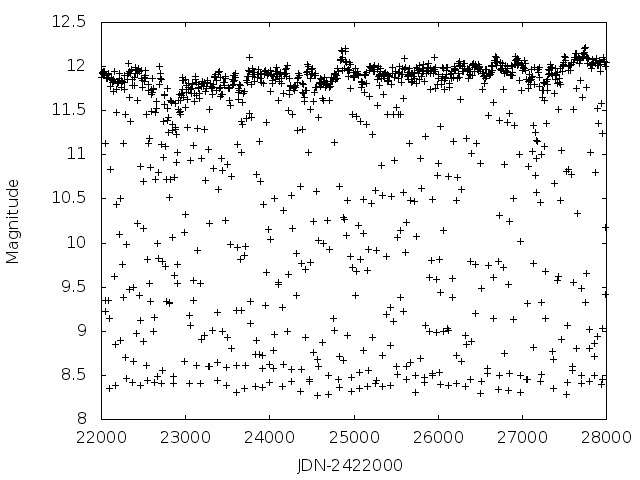}
\caption{ Light curve of variable star SS Cygni binned over 5 days}\label{fig:sscyglc}
\end{figure}
\subsection {Variable Star SS Cygni}

\begin{figure}[h!]
\centering
\plotone{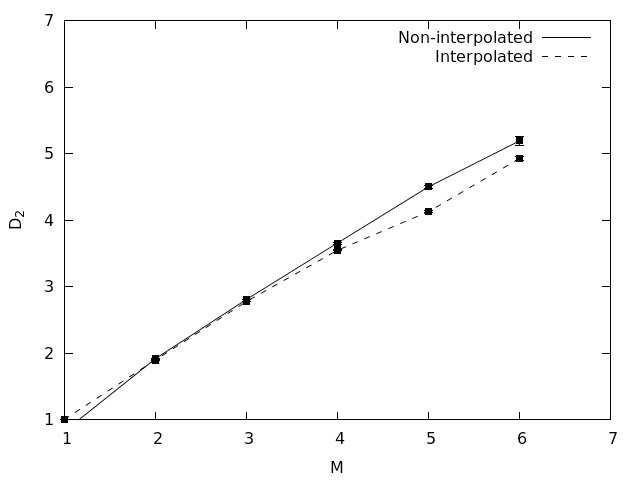}
\plotone{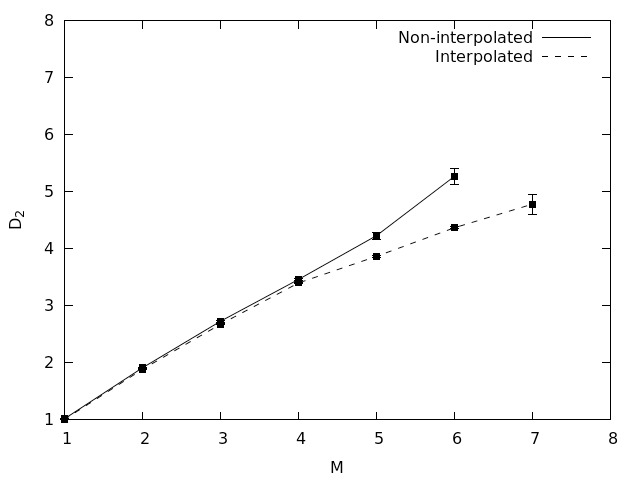}
\caption{ Correlation dimension D$_{2}$ vs embedding dimension, M, for variable star SS Cygni for the interpolated and non interpolated cases. While both curves do not saturate for case (a) for bins of 2 days, it saturates for the interpolated case in case (b) for bins of 5 days}\label{fig:sscyg}
\end{figure}

To check the effects of interpolation on a real data, we consider the data of SS Cygni which is a cataclysmic variable star, of the dwarf nova class. The light curve of this star has been known not to possess any chaotic dynamics, though it was a subject of debate in the early 1990s \citep{Hem90,Can92}. We first start with the AAVSO data set for this star from 1950 to 2014. The delay time,$\tau$, for this data set is found to be 10 days. Binning the data-set to 2 days yields a value of m$_{p}$ $\approx$ 10$\tau$ and m$_{s}$ $\approx$ 0.5$\tau$. The D$_{2}$ vs M graph does not saturate, even though the values of m$_p$ and m$_s$ do not fall into the identified critical range. A section of this light curve with binning to 5 days is shown in Fig.\ref{fig:sscyglc}. For this  m$_{p}$ changes to $\approx$50$\tau$ and the m$_{s}$ to $\approx$1$\tau$. The D$_{2}$ vs M graph still does not saturate as is expected from our earlier analysis. To demonstrate the effects of smoothing techniques, we interpolate using a cubic spline interpolation through both data sets and compute the D$_{2}^{sat}$ values using the smoothed data-sets. While the D$_{2}$ vs M curve does not saturate for the 2 day binned dataset, the 5 day binned and smoothed data set, gives the D$_{2}$ value of 4.139$\pm$0.05. We feel this illustrates the case where chaos is artificially introduced into a non-chaotic system due to interpolation. The D$_2$ vs M graphs for all cases are shown in Fig.\ref{fig:sscyg}.
\section{Conclusion} 
We analyze the variation in the value of computed correlation dimension D$_{2}$, due to presence of gaps in the observed or available data. For data from two standard chaotic systems, the R{\"o}ssler and the Lorenz, gaps are introduced artificially with Gaussian distributions for size and frequency of occurrence of gaps. We find with increasing value of mean position , m$_{p}$, the value of the correlation dimension falls to a constant value. This is because at large values of m$_{p}$, the bands of data between the gaps can yield values of correlation dimension which agree with each other. With increasing size of gaps, m$_{s}$ the value of D$_{2}$ first increases reaches a maximum at around m$_{s}\sim\tau$ and then falls to a constant value at higher values of m$_{s}$. This must be due to the fact that since the algorithm creates vectors sampled at every $\tau$, absence of a chunk of data of about the same size, can lead to a huge miscalculation of the correlation dimension. 

From the above analysis we can define an acceptable region of gap parameters where correlation dimension has a value comparable to the original value. Moreover we find that the value of embedding dimension for data with gaps is generally higher than the value from that without gaps. Hence we caution against using the value of embedding dimension while conducting further analysis like calculation of Lyapunov exponents, modeling equations for the system etc. The analysis using the surrogates for the Lorenz data indicates that even in cases where the D$_2$ values deviate away from the actual value due to presence of gaps, it still behaves differently from noisy data. 

In any real world time series, data often comes with total uneven or irregular sampling. With suitable binning such data can be converted to data with gaps of the type discussed here. Then the distribution of the total number of gaps and the size or extend of gaps can be determined. The means of both the distribution should help us to check whether they fall into the acceptable or unacceptable range of values for correlation dimension. Our main conclusions are validated for real world time series using observational data of variable stars, R Sct, U Mon and SU Tau. The chaotic nature underlying the variations in the light curve of variable star R Sct has been reported earlier. In case the other two stars, U Mon. and SU Tau, our results indicate that the saturation  of D$_2$ values, gives reliable conclusions regarding their fractal nature, since the gap distribution in both cases are outside the critical region.

We notice in our analysis, that a suitable choice of binning can improve the situation by causing the mean values to fall outside the critical region. In general, larger bins tend to average over small gaps, causing m$_s$ and m$_p$ to go up. This is exemplified in the case of U Mon, where one of the bins fall into the critical range for m$_{s}$ and does not yield a value for D$_{2}^{sat}$, whereas the other bin gives a saturated value. However there is often a trade-off between binning and retaining the fine structure of the data. In the case of the light curves we have considered, binning also tends to reduce the noise by averaging over the data. However while binning reduces the extend of uneven sampling, it also decreases the total number of available data points. Further large sized bins would tend to distort the fine structure of the data, which might be relevant for understanding the underlying dynamics of the system. From the analysis of the non chaotic light curve of variable star SS Cygni, we find that signatures of chaos can be introduced due to interpolation or smoothing of data, as previously suggested\citep{Gra86}. Therefore in any real world data, smoothing techniques like interpolation should be applied with caution. Moreover in many cases, depending on the profile of the gaps in the data, such techniques may be unnecessary for detecting underlying nonlinearity using correlation dimension. 

\acknowledgments We acknowledge with thanks the variable star observations from the AAVSO International Database contributed by observers worldwide and used in this research.

\end{document}